\documentclass[10pt,prd,longbibliography,superscriptaddress]{revtex4-2}

\usepackage[T1]{fontenc}
\usepackage[utf8]{inputenc}
\usepackage[english]{babel}
\usepackage{amsfonts,amsbsy,amssymb,amsmath,graphicx,float} 
\usepackage{color}

\graphicspath{{figures/}}

\usepackage[linktoc=all,colorlinks=true,citecolor=blue,linkcolor=blue]{hyperref}
\begin{document}

\markboth{Garc{\'i}a-S{\'a}nchez, Mancho, Agaoglou \& Wiggins}{New links between invariant dynamical structures and uncertainty quantification.}

\title{New links between invariant dynamical structures and uncertainty quantification.}

\author{Guillermo Garc{\'i}a-S{\'a}nchez}
\address{Instituto de Ciencias Matematicas,CSIC, C/Nicolas Cabrera 15, Campus Cantoblanco, Madrid, 28049, Spain}
\address{Escuela T{\'e}cnica Superior de Ingenieros de Telecomunicación, Universidad Polit{\'e}cnica de Madrid, Av. Complutense, 30, Madrid,28040,Spain}
\author{Ana Maria Mancho}
\address{Instituto de Ciencias Matematicas,CSIC, C/Nicolas Cabrera 15, Campus Cantoblanco, Madrid, 28049, Spain}
\author{Makrina Agaoglou}
\address{Departamento de Matem{\'a}tica Aplicada a la Ingenier{\'i}a Industrial, Escuela T{\'e}cnica Superior de Ingenieros Industriales, Universidad Polit{\'e}cnica de Madrid, c/ Jos{\'e} Guti{\'e}rrez Abascal, 2, Madrid, 28006,Spain}
\author{Stephen Wiggins}
\address{School of Mathematics, University of Bristol, Fry Building, Woodland Road, Bristol, BS8 1UG, United Kingdom.}      
\address{Department of Mathematics, United States Naval Academy, Chauvenet Hall, 572C Holloway Road, Annapolis, MD 21402-5002}

\begin{abstract}
This paper proposes a new uncertainty measure, appropriate for quantifying the performance of transport models in assessing the origin or source of a given observation. It is found that in a neighbourhood of the observation the proposed uncertainty measure  is related to the invariant dynamical structures of the model.
The paper illustrates the implementation of the proposed definition to quantify the performance of ocean data sets in the context of a real oil spill event in the Eastern Mediterranean in 2021.
    \end{abstract}

\maketitle

\section{Introduction}
Models in real-world applications are subjected to uncertainty 
in their results, and this is particularly notorious in models that describe the ocean or the atmosphere state. In these geophysical contexts, the availability of tools that  quantify the uncertainty of model outputs is very much desired as this allows the evaluation of their reliability and the assessment of their use in real applications. Uncertainty quantification is a very broad question and it has been addressed from different perspectives. One approach to this problem involves performing a sensitivity analysis, which determines what inputs in the model, which typically is a partial differential equation, affect the outputs the most. A related issue, referred to as  forward propagation of uncertainty, consists of examining how uncertainty in the model parameter inputs affects uncertainty in the model outputs. In this framework it may happen that experimental data are available. In this case, it is possible  to infer the set of parameters that better fit the outputs for the proposed model. This setting can be precisely formulated as a Bayesian inverse problem, which for a given observation, and an assumed noise model, determines the parameters that are most likely to have produced the data. Typically, Bayesian analyses provide the framework for many inverse uncertainty quantification applications \cite{wu2018inverse, domitr2022comparison}. These
methods employ  Bayesian inference theory along with exploring the posterior probability density function (PDF) by Markov Chain Monte Carlo (MCMC) sampling. Other methods include statistical methods based on maximum likelihood estimation (MLE) \cite{de1996determination} and data assimilation methods. Data assimilation methods combine experimental observations with code predictions and their respective errors to provide an improved estimate of the system state and of the associated uncertainty \cite{petruzzi2019casualidad}.
All of these approaches assume that the model is known, with undetermined parameters, and perform adaptations on  the model inputs. 
However, there is a different type of problem for which  only an approximate model is available and its exact expression is unknown. We delve into an explanation of such problems.


Recently, Garc\'ia-S\'anchez et al. \cite{guillermo2022} have proposed to quantify uncertainty associated with transport by currents in 
settings in which 
the exact model  connecting two successive observations (or the currents producing the observed transport) was not known. They have quantified uncertainty  
motivated by the need to judge the performance of ocean model outputs to describe  oil spill events such as that described in \cite{garciasanchez2020}. In this reported event an identified spill occurs at a certain time and it is required  to describe its evolution to determine if it will affect critical areas, and there is a need to assess ocean models to this end.  
Garc\'ia-S\'anchez et al. \cite{guillermo2022}  approach evaluates the reliability of  the ocean currents with respect to the transport that they produce. Indeed, 
transport in the  ocean surface is produced by fluid  parcels that follow trajectories ${\bf x}(t)$ that evolve according to the dynamical system:
\begin{equation}
    \frac{d{\bf x}}{dt} = {\bf v}({\bf x},t),
    \label{v(t)}
\end{equation}
This system  is a nonlinear non-autonomous dynamical system in which there are uncertainties in the velocities ${\bf v}({\bf x},t)$, because  they are  not exactly known given that they come from solving highly sensitive partial differential equations. This is what we mean by {\em not knowing} the exact problem: the right-hand side of Equation \eqref{v(t)} is only an approximation. Typically, in other problems where uncertainty has been studied, the system obeys an explicitly known partial differential equation where the unknowns are not within the model itself but rather in adjustable parameters of the equation or in the observations. However, our case is of a different nature because the right-hand side of Eq. \eqref{v(t)} is a key part of the equation and not exactly known. The uncertainties in the velocities ${\bf v}({\bf x},t)$ produce uncertainties in the solutions ${\bf x}(t)$.
Garc\'ia-S\'anchez et al. \cite{guillermo2022}  quantified this type of  uncertainty by means of error measures in 
settings  where  the initial observation was known, ${\bf x_0}$,  and a target state at a later final time,  ${\bf x^*}$. This approach was adequate to quantify the suitability of the model given in Eq. \eqref{v(t)} to represent this sequence of observations   in forward time.  In doing so they found  links between the uncertainty and the stable invariant manifolds of hyperbolic trajectories present in the system  \eqref{v(t)}. 

This paper aims to complete the described picture as follows.
While results in  \cite{guillermo2022} consider the uncertainty of a model based on ocean currents in order to describe  the final fate of an observation as it evolves {\em forward} in time, the new setting considers the uncertainty of a similar model in order to identify the original location of an observation in a {\em past} time, once  the  position, ${\bf x_1}$,  of this observation is known at the present time. In a nutshell, previous results were about: where do trajectories end up? How could it be measured if they are consistently described by a model  constructed from approximate velocities? Now the results we desire are about: where do trajectories come from? how could it be measured if their origin is  consistently described by a model that is constructed from approximate velocities? In this context, this paper provides a new approach to measuring uncertainty using errors adequate for the backward time setting. We find that similar to what was found in Garc\'ia-S\'anchez et al. \cite{guillermo2022}, this uncertainty has a {\em structure} in a neighborhood of the observation  ${\bf x_1}$, and this structure is related to the unstable invariant manifolds of the  hyperbolic trajectories present in the vector field of equation \eqref{v(t)}.



Once this new type of uncertainty is introduced, we proceed to assess ocean data capabilities in  real events that demand this type of perspective.  In particular, a recent contamination event that occurred in 2021 in the Eastern Mediterranean is considered. Specifically in this event, thoroughly described in \cite{garciasanchez2022}, several beaches were affected by a spill of unknown origin.  In contrast to the results by \cite{guillermo2022},  now observations are done at the arrival point and the identification of their source is the goal. Equipped with the new tool we describe in detail its use to quantify the performance of different ocean data sets for describing observations adequately.    
There are other attempts to identify the best  dataset for predicting the drift of oil spills \cite{brushett2011evaluation, zhang2020evaluation}. However, their approach consists of using uncertainty estimators in forward time.

In summary, the goal of this article is to propose a new uncertainty measure appropriate for quantifying the suitability of a model to describe the origin   of trajectories. It is found that the considered uncertainty has a structure correlated  with the unstable manifolds of the hyperbolic trajectories of the model. The capacity of this new measure is exploited for characterizing  the performance of ocean data sets in real events. Accordingly, the structure of this article is as follows. Section 2  provides a new definition for uncertainty quantification suitable to describe the new setting and,  in a simple example, discusses  links with unstable manifolds of hyperbolic trajectories. Section 3  presents the application of this tool to the mentioned oil spill event that affected the Eastern Mediterranean, and shows how it helps to discriminate between two data sets, and supports the identification of the most suitable one for searching the spill source. In Section 4, the paper concludes with a discussion of the implications and potential applications of this approach.

\begin{figure}[htb!]
  \begin{center}
\includegraphics[scale=0.5]{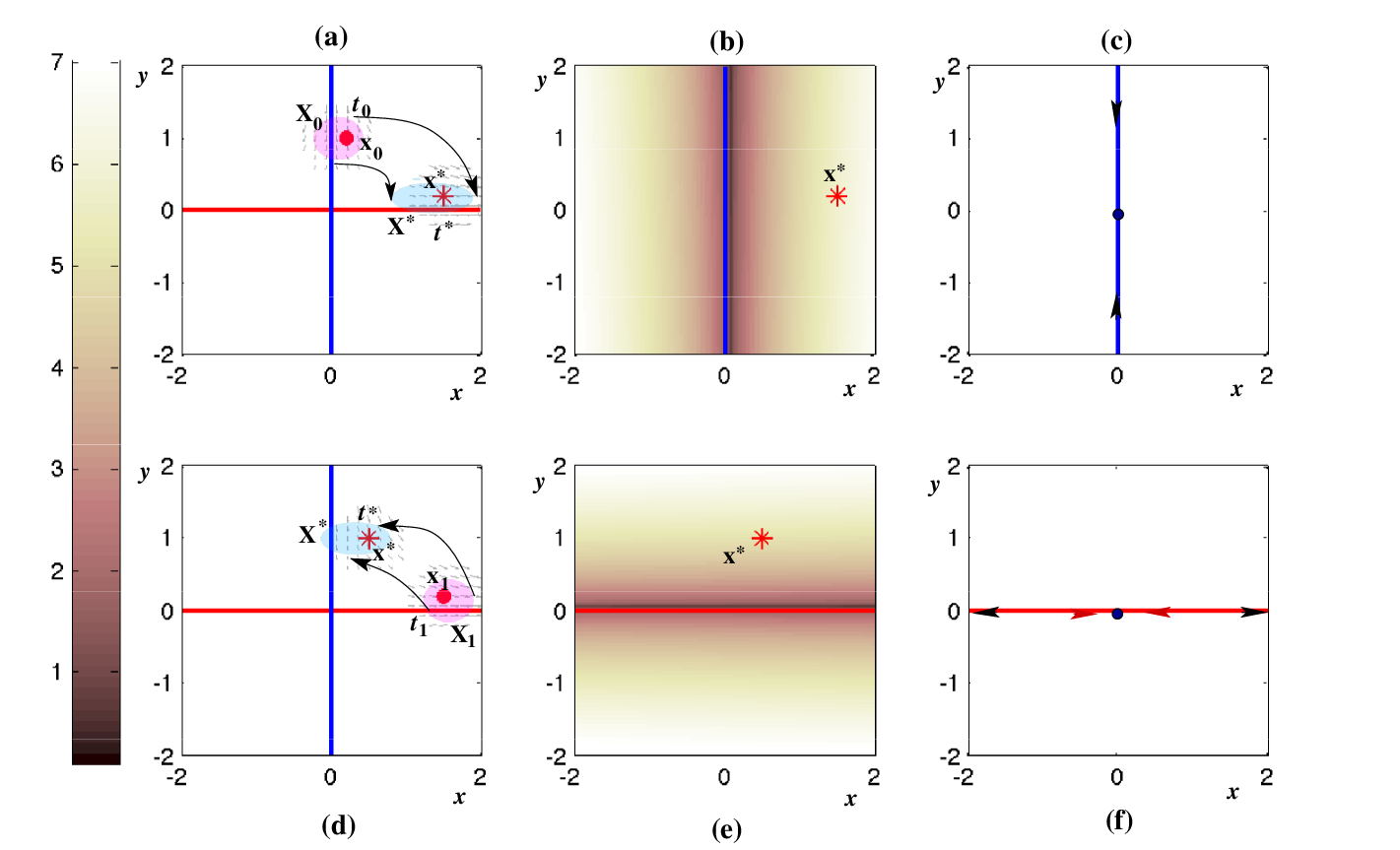} 
  \end{center}
 \caption{Panels a), b) and c) display a graphical representation of two sequential observations and their forward time evolution. a) The initial observation at time $t_0$ is expressed by the red circle with initial condition ${\bf x_0}$. The final observed state  
 ${\bf x^*}$ at time $t^*$ is referred to as the  "target". In general, the  evolution law linking these observations is unknown, but  in our setting, it is approximated by Eq. \eqref{eq:saddle}, a system with a hyperbolic fixed point at the origin with  stable and unstable manifolds in blue and red, respectively. 
 The forward evolution  of ${\bf X_0}$ (the pink blob) to ${\bf X^*} $  (the light blue blob) according to this model is illustrated; b) a representation of $L_{UQ}$, in a domain beyond the neighbourhood $X_0$ for $\tau=3$, where singular features  are identified aligned with the stable manifold (in blue); c) a representation of the stable manifold. Black arrows point out the forward time evolution for initial condition on the stable manifold   towards the fixed point at the origin; Panels d), e) and f) display a  graphical representation of two sequential observations and their backward time evolution, d) the final observation at time $t_1$ is expressed with the red circle at the position ${\bf x_1}$.  Although both the origin and the model of this observation are unknown,  its evolution is  approached by  Eq. \eqref{eq:saddle}, a system with a hyperbolic fixed point at the origin with  stable and unstable manifolds in blue and red respectively. 
 The backward evolution  of  ${\bf X_1}$ (the pink blob) to ${\bf X^*} $ (the light blue blob) according to this model is illustrated; e) a representation of $L_{BUQ}$, in a domain beyond the neighborhood $X_1$ for $\tau=3$, where singular features  are identified aligned with the unstable manifold (in red); f) a representation of the unstable manifold. Red arrows point out the evolution in backward time  towards the fixed point for initial conditions on the unstable manifold.  Black arrows point out the evolution forward time  moving away from the fixed point. The color bar at the top indicates a color scale for $L_{UQ}$ and $L_{BUQ}$ in panels b) and e).}
  \label{fig:ex}
\end{figure}

\section{Uncertainty Quantification in Backward Time}


This section is focused on proposing a new definition of uncertainty quantification to describe the uncertainty in the origin or source of certain observations based on predictions by a model constructed from velocities. This section is limited to present the definition and to discuss   its implementation in one simple example.  This simplified scenario enables us to establish connections between the new definition of uncertainty and the unstable manifolds of hyperbolic trajectories. The objective, however, is to apply this definition to distinguish the most appropriate velocity field among various options, the one that best matches the observations.

We start by recalling results reported in \cite{guillermo2022} in the context of the simple example illustrated in
Figure \ref{fig:ex}. Panel a) within this figure shows an initial observation, ${\bf x}_0$, at time $t_0$ marked with a red circle. At a later time, $t^*$, this observation is at a target position, ${\bf x}^*$,  marked with a red asterisk.  This evolution is assumed to be following the model provided by  Eq. \eqref{eq:saddle}, with a vector field in the background, which is represented at time $t_0$, in the neighborhood of ${\bf x}_0$ and at time $t^*$ in the neighborhood of ${\bf x}^*$. This vector field is chosen to be  stationary, therefore its representation at later times, $t^*$, is the same as at $t_0$. However, in general, this vector field will be time-dependent and therefore its representation at $t^*$ is different from that at $t_0$. The explicit expression for this vector field is:
\begin{equation}\label{eq:saddle}
    \begin{cases} 
    \frac{dx}{dt}  = x, \\ 
    \frac{dy}{dt}  = - y,
     \end{cases} 
\end{equation}
This system has a stable manifold aligned with the vertical axis at $x=0$ (blue line in Figure \ref{fig:ex}) and an unstable manifold aligned with the horizontal axis at $y=0$ (red line in figure \ref{fig:ex}).

Garc\'ia-S\'anchez et al. \cite{guillermo2022} proposed several expressions to measure the error in a neighborhood ${\bf X}_0$ of ${\bf x}_0$. Among those we consider this one:
\begin{equation}
    L_{UQ}({\bf X}_0, t_0, \tau, p) = \sum_{i=1}^n\left|x_i(t_o+\tau)-x_{i}^*\right|^p , \ p\le 1,   \ {\bf X}_0\in \mathbb{R}^n,  \tau>0.
    \label{eq:UQ1}
\end{equation}
Here,  ${\bf X}_0$ represents a set of initial conditions whose coordinates  $(x_1,x_2,..,x_n)$, belong to a $n$  dimensional  system \eqref{v(t)} and are in the neighbourhood of the observation ${\bf x}_0$. The coordinates of the target, towards which the system is assumed to evolve in a time interval $\tau$, are $(x_{1}^*,x_2^*,..,x_n^*)$. Panel a) in Figure \ref{fig:ex} illustrates the evolution from the set ${\bf X}_0$ (the pink blob) to ${\bf X}^*$ (the light blue blob).
Panel b)    
displays  $L_{UQ}$ evaluated in the whole domain. This representation expresses  how much  uncertainty would be associated with the system \eqref{eq:saddle} beyond the neighborhood ${\bf X}_0$,  assuming that initial observations  start at different points of the representation domain and that they evolve towards the assumed observed target position at the red asterisk in  $\tau=3$. It is clear from the representation, that $L_{UQ}$ presents a structure  that consists of singular features aligned with the stable manifold of the hyperbolic fixed point present in the vector field, as discussed and proven in \cite{guillermo2022}.

\begin{figure}[htb!]
  \begin{center}
 \includegraphics[scale=0.6]{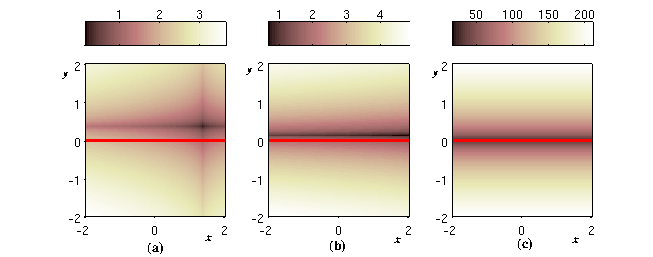} 
  \end{center}
 \caption{A representation in the plane of $L_{BUQ}$  for the dynamical system \eqref{eq:saddle} with target $x^*=(0.5,1 )$. a) $\tau=1$; b) $\tau=2$; c) $\tau=15$. The red line represents the position of the unstable manifold.}
  \label{fig:ex2}
\end{figure}

In this paper, our interest is in quantifying the uncertainty of a model for identifying a target source  ${\bf x}^*$,  which is consistent with a later observation ${\bf x}_1$ at time $t_1$. In this  way,  the target source, ${\bf x}^*$, is located at an earlier time $t^*=t_1-\tau$. Similarly to the previous definition, for this setting, we propose the uncertainty to be:
\begin{equation}
    L_{BUQ}({\bf X}_1, t_1, \tau, p) = \sum_{i=1}^n\left|x_i(t_1-\tau)-x_{i}^*\right|^p , \ p\le 1,   \ {\bf X}_0\in \mathbb{R}^n,  \tau>0.
    \label{eq:BUQ}
\end{equation}
where now  ${\bf X}_1$ represents a neighbourhood around the observation ${\bf x}_1=(x_1,x_2,..,x_n)$ at time $t_1$  being $n$  the dimension of the system \eqref{v(t)}, which in our specific examples is $n=2$. The coordinates of the target are $(x_{1}^*,x_2^*)$, and uncertainty is provided in terms of a distance metric  between the target and the {\em backward} evolution of points near to ${\bf x}_1$ for a period $\tau$. 

Panel d) in Figure \ref{fig:ex}  illustrates the backward in time evolution of ${\bf X}_1$ (the pink blob) to ${\bf X}^*$ (the light blue blob) 
  In the background, the panel also illustrates  the vector field  in a neighborhood of ${\bf x}_1$ at time $t_1$, and in the neighborhood of ${\bf x}^*$ at time $t^*$. Given that this vector field is stationary,  its representation at an earlier time, $t^*=t_1-\tau$, will be the same as at $t_1$, although  in general this vector field is time-dependent and therefore its representation at $t^*$ is different from that at $t_1$. 

Panel e)    
displays in the whole domain, that is beyond the neighborhood  ${\bf X}_1$, $L_{BUQ}$ evaluated for $p=0.5$,  assuming $\tau=3$. This representation expresses  how much uncertainty 
is associated to the system \eqref{eq:saddle}, assuming that  observations  ${\bf x}_1$ could be   anywhere in the domain and that they have  evolved there from the target position ${\bf x}^*$ at $t^*=t_1-\tau$. It is clear from the representation that  $L_{BUQ}$ presents a structure  that consists of singular features aligned with the unstable manifold of the hyperbolic fixed point present in the vector field. A formal proof of this observation is given in the appendix. In the proof it is shown how  a sufficiently large $\tau$ is required for the singular structures obtained from Eq. \eqref{eq:BUQ} to become aligned with the unstable manifolds. This is more clearly visualized in Figure \ref{fig:ex2}. Panel a) shows the evaluation of $L_{BUQ}$ for $\tau=1$, where clearly the alignment with the unstable manifold is not yet achieved. Panels b) and c) confirm the convergence of the alignment towards the unstable manifold for, respectively, $\tau=2$ and $\tau=15$.
This result completes the link between invariant dynamical structures and uncertainty quantification.  

Panels c) and f) in Figure \ref{fig:ex} graphically illustrate why the forward time uncertainty for system \eqref{eq:saddle}   is related to the stable manifold of the hyperbolic fixed point, while the backward time uncertainty  is linked to the unstable manifold of that fixed point.  Indeed, the forward time integration for initial conditions that stay on the stable manifold, for a sufficiently large $\tau$, approach the fixed  point at the origin, maintaining a fixed distance to the target, while all other initial conditions in the plane will increase their distance to the target for increasing $\tau$. This is the origin of the singular features aligned with the stable manifold in forward time. 
Analogously, the backward time integration of initial conditions that stay on the unstable manifold, for a sufficiently large $\tau$, approaches the fixed  point at the origin, maintaining a fixed distance to the target. All other initial conditions in the plane will increase their distance to the target for increasing backward  integration. This is the origin of the singular features aligned with the unstable manifold in backward time. Analogously this explanation also illustrates why both   $L_{UQ}$ and  $L_{BUQ}$ take minimum values along the unstable manifold. Indeed,  they express the finite distance between the fixed point and the target, while the value of $L_{UQ}$ and  $L_{BUQ}$ diverge for growing $\tau$ for all other points in the plane. 

\begin{figure}[htb!]
  \begin{center}
 \includegraphics[scale=0.6]{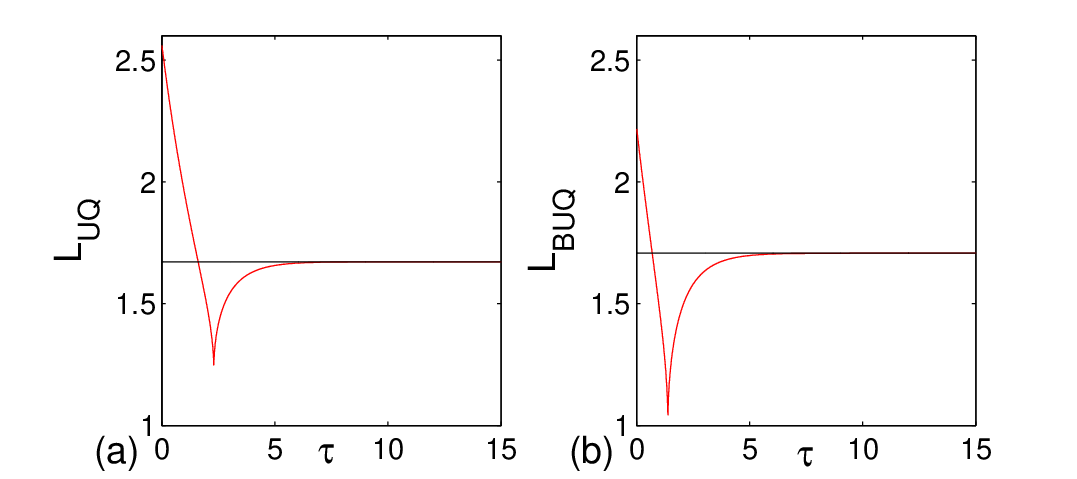} 
  \end{center}
  \caption{For the system given by  Eq. \eqref{eq:saddle}, a) evolution of  $L_{UQ}$ versus $\tau$ on the fixed point  (black line) and on a point on the stable manifold (red line) with target ${\bf x}^*=(1.5,0.2)$; b) evolution of  $L_{BUQ}$  versus $\tau$ on the fixed point (black line) and on a point on the unstable manifold (red line) with target ${\bf x}^*=(0.5,1)$. }
  \label{fig:tau}
\end{figure}

Figure \ref{fig:tau} illustrates further these ideas. Panel a)   represents the evolution versus $\tau$ of $L_{UQ}$ at points: (0,2) in red -- which is on the stable manifold-- and   (0,0) in black --the hyperbolic fixed point--  as a function of $\tau$, both with a target ${\bf x}^*=(1.5,0.2)$ and $p=0.5$.  
There is a transient behavior in $\tau$ for the point on the stable manifold, while in the limit of large $\tau$ both values converge to $|1.5|^{0.5}+|0.2|^{0.5}\sim1.672$, which is the distance, in the defined metric, between the fixed point and the target. Similarly, panel b) represents  the evolution of $L_{BUQ}$ at points: (2, 0) in red --which is on the unstable manifold-- and (0,0) in black --the hyperbolic fixed point-- as a function of $\tau$, both with a target ${\bf x}^*=(0.5,1)$ and $p=0.5$. In the limit of large $\tau$, both converge to the constant $\sim 1.71$.

\begin{figure}[htb!]
  \begin{center}
 \includegraphics[scale=0.4]{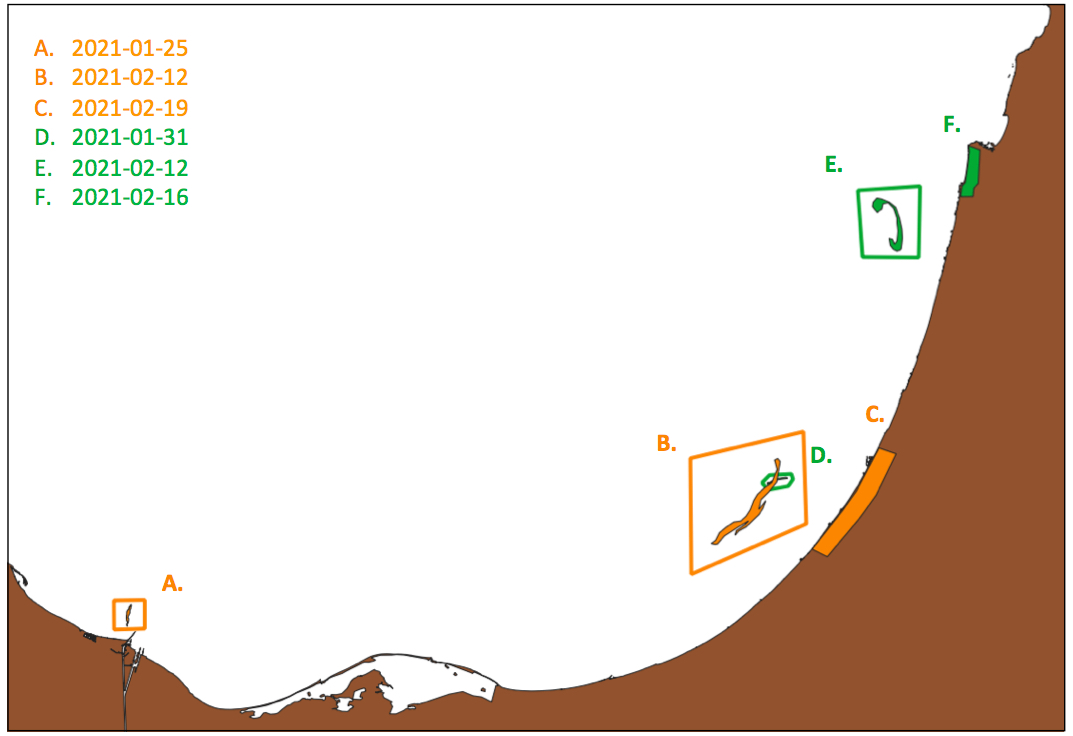} 
  \end{center}
  \caption{A graphical representation of the spills   observed along the coastline of the Eastern Mediterranean and satellite observations matching the sources. }
  \label{fig:med}
\end{figure}




\section{Results}

The backward-time uncertainty definition provided in Eq. \eqref{eq:BUQ}, or its analog for the Euclidean distance given by:  
\begin{equation}
    L_{BUQ}({\bf X}_1, t_1, \tau) = \left[\sum_{i=1}^n\left|x_i(t_1-\tau)-x_{i}^*\right|^2 \right]^{\frac{1}{2}},  \ {\bf X}_1\in \mathbb{R}^n, \tau> 0
    \label{eq:BUQ2}
\end{equation}
are very suitable for assessing the performance of ocean models for describing events where the source of a certain observable,  for instance, a contaminant, is unknown.   This section describes how  to implement the use of the definitions given in \eqref{eq:BUQ} or \eqref{eq:BUQ2} to assess and determine
the model that better achieves the observations in a recent oil spill of this type. This event is of a different nature to those  described in \cite{garciasanchez2020}, where after a reported oil spill accident  the interest was in describing and controlling its evolution in order to minimize its impact. In contrast,  now we are interested in assessing models that provide information about when and where did   the spill originate.  
Good models in the setting described in Figure  \ref{fig:ex}d) are expected to have small distances between ${\bf X^*}$ (the blue domain) and ${\bf x^*}$ (the target) and we will exploit this idea to measure the quality of available ocean data sets.

We describe next the specific event to be considered under this perspective. At the beginning of  2021, the coastline of several Middle Eastern countries in the Eastern Mediterranean was affected  by the presence of  oil from an unknown source(s) \cite{WP}. No accidents were reported for these spills previous to their arrival at the coast, and therefore there was  no hint of their origin.   Israeli authorities
estimated that more than 1000 tons of tar \cite{haar} landed along 180 km  of the Israeli and Lebanese shoreline \cite{jp1,jp2, jp3, ay1,rut,ata} in mid-February.  They combined the use of satellite images with simulations by Garcia-S\'anchez et al. \cite{garciasanchez2022},   which provided possible sources for this spill. Their work studied this event both from the remote sensing and modeling perspectives.  Figure \ref{fig:med} illustrates some affected geographical areas. In this figure,  orange and green colors along the coastline mark the oil arrivals.
The systematic search of satellite images identified potential sources of these observations and a careful examination concluded that those sources surrounded by green and orange polygons, which correspond to satellite observations,  were located in position and time in a manner consistent with the position and dates of the oil that subsequently arrived at the coast. According to \cite{garciasanchez2022}, orange polygons labeled as A and B, are  successive earlier locations  to the orange arrival labeled as C. Analogously green polygons labeled as D and E, correspond to the green arrival labeled as F. The successive dates for each observation are given in the figure. 

Simulations in \cite{garciasanchez2022} demonstrated that the spill was closely following fluid parcel  trajectories ${\bf x}(t)$ that obeyed the equation:
\begin{equation}
    \begin{cases}
    \dfrac{d \lambda}{d t} = \dfrac{u(\lambda,\phi,t)}{R\cos \phi} \\[.3cm]
    \dfrac{d \phi}{d t} = \dfrac{v(\lambda,\phi,t)}{R}
    \end{cases}\label{v(t)s}
\end{equation}
where  the position  of the fluid parcel at the ocean surface is given in longitude ($\lambda$) and latitude ($\phi$), and $R$ is the Earth's radius.  The system \eqref{v(t)s} is the particular form for our study of the general expression given by Eq. \eqref{v(t)}, where  ${\bf x}=(\lambda,\phi)$  and   its right-hand side represent the two components of the vector field ${\bf v}$, which are determined by the zonal ($u$) and meridional ($v$) velocities.  These velocity components  are obtained as data sets from the Copernicus Marine Monitoring Environmental Service (CMEMS). 

Equation \eqref{v(t)s} is a nonlinear non-autonomous dynamical system in 2D and it is required for implementing the backward evolution of fluid parcels in expression \eqref{eq:BUQ2}. $L_{BUQ}$ in Eq. \eqref{eq:BUQ2} will have different evaluations for different CMEMS data sets. The goal of this section is to use these outputs to discriminate  models. 
CMEMS provides two data sets describing currents in the Mediterranean Sea. One is the operational Mercator global ocean analysis and forecast system that 
provides 10 days of 3D global ocean forecasts updated daily. 
This product includes daily and monthly mean fields of variables such as temperature, salinity, currents, sea level, mixed layer depth, and ice parameters from the top to the bottom of the global ocean. It also includes hourly mean surface fields for sea level height, temperature, and currents. The global ocean output files are displayed with a 1/12 degree horizontal resolution with an equirectangular  longitude/latitude  projection and 50 vertical levels  ranging from 0 to 5500 meters. 

The second data set  is the physical component of the Mediterranean Forecasting System (Med-Currents), which is a coupled hydrodynamic-wave model, including tides, implemented over the whole Mediterranean Basin, with a higher resolution than the global model. Indeed, its horizontal grid resolution is 1/24 degree 
with 141 unevenly spaced vertical levels.  
The model solutions are corrected with a
variational data assimilation scheme (3DVAR) of vertical profiles of temperature and salinity along the satellite track that provides observations of sea level anomalies (SLA).


Given that both models provide information for the same area, a natural question is to determine whether one is more suitable than the other for describing transport in the type of event described above. Models such as that given in Eq. \eqref{v(t)s}, possess a transport signature based on invariant manifolds associated with hyperbolic trajectories, which following Poincar{\'e}’s idea, constitute geometrical structures on the ocean surface that organize particles schematically by regions corresponding to qualitatively different types of trajectories. These geometrical features allow a more robust analysis of the  transport capacity of ocean currents than that based on individual trajectories. 
Indeed, there exist mathematical results that discuss the persistence  of these geometrical structures  in the presence of small perturbations, while individual trajectories may be  affected much more drastically by perturbations \cite{gg2018,haller2002}.
In the past these features  have been used to assess the transport performance of data sets \citep{kuznetsov2002,shadden2009,haza2007,haza2010,mendoza2014lagrangian,gg2015}. In the context of geophysical flows, these geometrical structures  are referred to as Lagrangian Coherent Structures (LCSs). The use of LCS allows  a {\em qualitative}  assessment of the performance of data sets, however, definitions such as that given in Eq. \eqref{eq:BUQ2} allow a  {\em quantitative} analysis that we will implement next in the context of the oil spill event described in \cite{garciasanchez2022}.

In this work, we compute LCS using Lagrangian Descriptors \cite{mendoza2010, mancho2013lagrangian}, a method  based on computing  the function defined as:
\begin{equation}
M(\mathbf{x}_{0},t_0,\tau) =  \int^{t_0+\tau}_{t_0-\tau}  ||\mathbf{v}(\mathbf{x}(\mathbf{x}_0,t), t)|| \; dt   , 
\label{eq:Mp_function}
\end{equation}
where  the   Euclidean norm, $||.||$,  of the vector field $\mathbf{v}$ is evaluated along fluid parcel trajectories $\mathbf{x}(\mathbf{x}_0,t)$ of  Eq. \eqref{v(t)s},  and therefore the integral in Eq. \eqref{eq:Mp_function} evaluates the arclength of this trajectory  on the latitude-longitude plane at the ocean surface. In expression \eqref{eq:Mp_function}, the forward integration from $t_0$ reveals by means of singular features the stable manifolds associated to hyperbolic trajectories, whereas the backward integration highlights their unstable manifolds.
This is explicitly expressed by splitting the integral \eqref{eq:Mp_function}  into two terms as follows:
\begin{equation}
M(\mathbf{x}_{0},t_0,\tau) = \int_{t_0-\tau}^{t_0} \mathbf{v}(\mathbf{x}(\mathbf{x}_0,t), t) \; dt   \;+ \int_{t_0}^{t_0+\tau}  \mathbf{v}(\mathbf{x}(\mathbf{x}_0,t), t) \; dt =    M^{(b)}(\mathbf{x}_{0},t_0,\tau)+ M^{(f)}(\mathbf{x}_{0},t_0,\tau)
\label{eq:split}
\end{equation}

\begin{figure}[htb!]
 \begin{center}
   a) \includegraphics[scale=0.4]{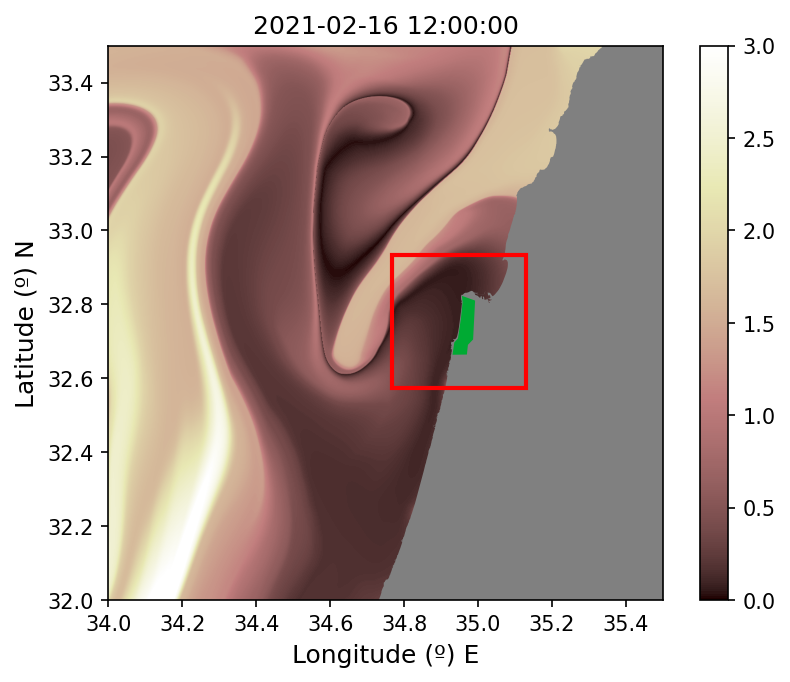} 
    b) \includegraphics[scale=0.4]{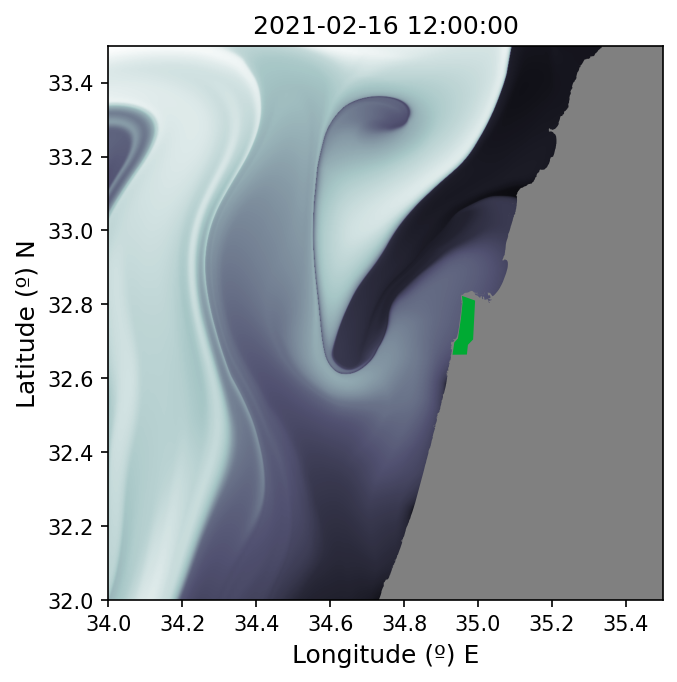} \\
	    c) \includegraphics[scale=0.4]{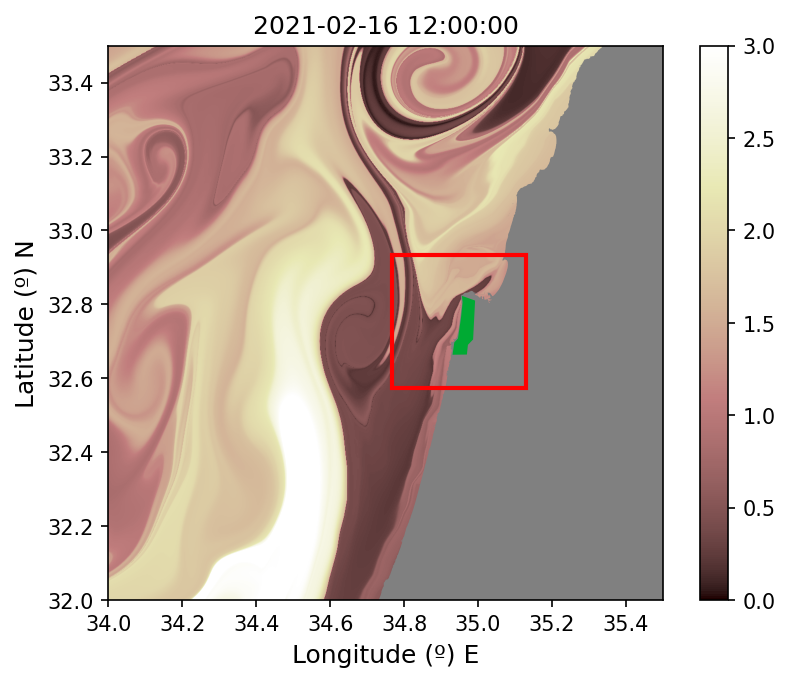} 
    d) \includegraphics[scale=0.4]{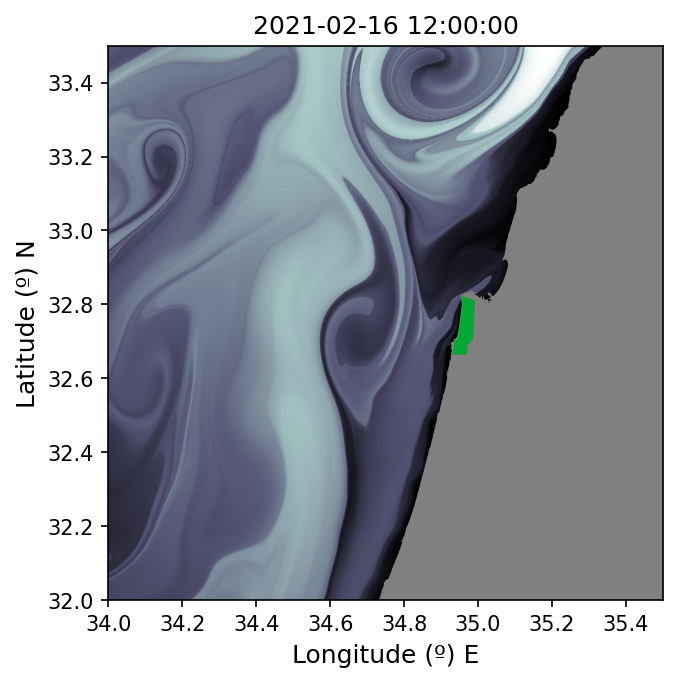}
\end{center}
   \caption{Evaluation on the 16th of February 2021 of  $L_{BUQ}$ using the target, ${\bf x^*}=(34.36^\circ \textrm{E}, 31.78^\circ \textrm{N})$ and of $M^{(b)}$ using $\tau=16$ days; a) $L_{BUQ}$ with the  CMEMS global product; b) $M^{(b)}$ with the  CMEMS global product; c) $L_{BUQ}$ with the  CMEMS Mediterranean product; d) $M^{(b)}$ with the  CMEMS Mediterranean product.}
	\label{fig:globmed1}
 \end{figure}
We explore next the implementation of the ideas discussed in the previous section, to evaluate the performance of the two CMEMS products in the context of this environmental damage and their links to the unstable manifolds of hyperbolic trajectories present in the system Eq. \eqref{v(t)s}. Following oil detections displayed in figure \ref{fig:med}, we start the discussion considering the first  arrival to the coast on the 16th of February 2021, which is the green  mark at the Israeli coast (F) and is considered to be  the observation ${\bf x_1}$. For it, we consider that the position (D) on the 31st of January 2021 is its backward time target ${\bf x^*}=(34.355734^\circ \textrm{E}, 31.777414^\circ \textrm{N})$. 
Figure \ref{fig:globmed1}a) shows the evaluation of $L_{BUQ}$ according to Eq. \eqref{eq:BUQ2} in a large area, considering that the eastward ($u$) and  northward ($v$) velocity components are provided by the CMEMS global model. It is remarkable that its structure presents important analogies with the $M^{(b)}$, visible in panel b), which has been computed for $\tau=16$ days. This image   highlights, by means of singular features, the unstable manifolds of hyperbolic trajectories present in the vector field supplied by  Eq. \eqref{v(t)s}. This result  confirms that the uncertainty associated with the model has a structure closely related to invariant dynamical structures. 
The red square in panel a) highlights a neighborhood of the observation whose origin is conjectured to be at target ${\bf x^*}$. This 
 domain ${\bf X_1}$  is used to characterize the performance of the model with respect to the hypothesis, assuming that the model is good if the target is consistent with arrivals in all this domain ${\bf X_1}$. The model is characterized through mean values of  $L_{BUQ}$ in this box. Panels c) and d) reproduce results for the CMEMS Mediterranean  product. 
It is again clear the close relationship between the structure of $L_{BUQ}$ and the unstable manifolds of hyperbolic trajectories present in   Eq. \eqref{v(t)s}. 

\begin{figure}[htb!]
 \begin{center}
   a) \includegraphics[scale=0.4]{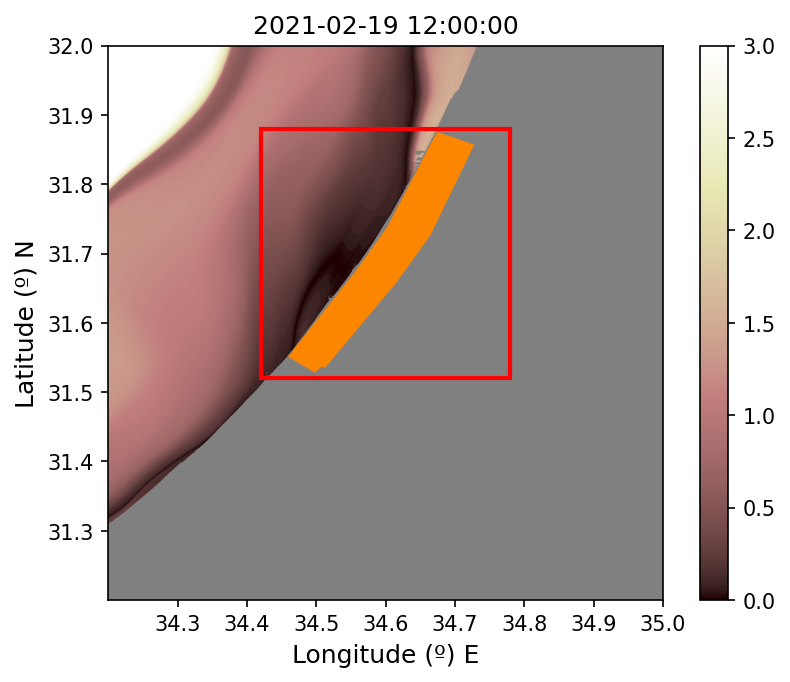} 
    b) \includegraphics[scale=0.4]{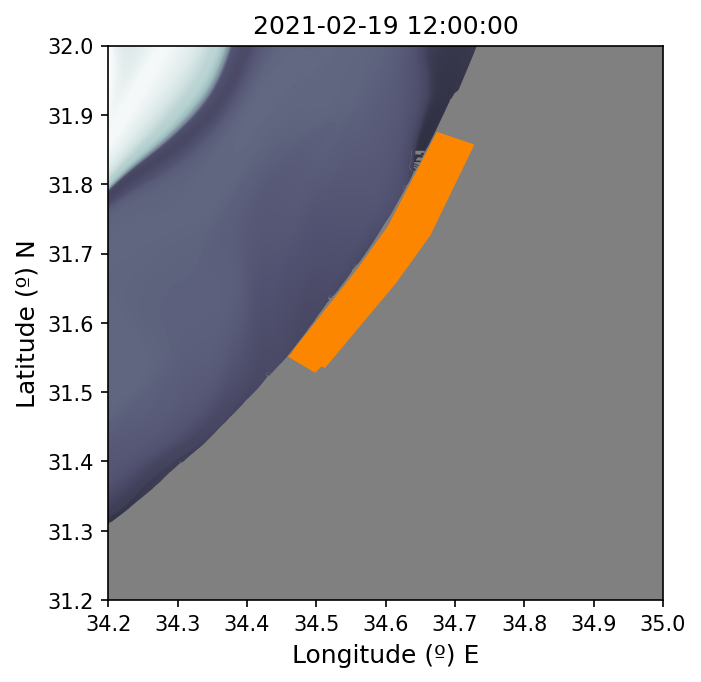} \\
   c) \includegraphics[scale=0.4]{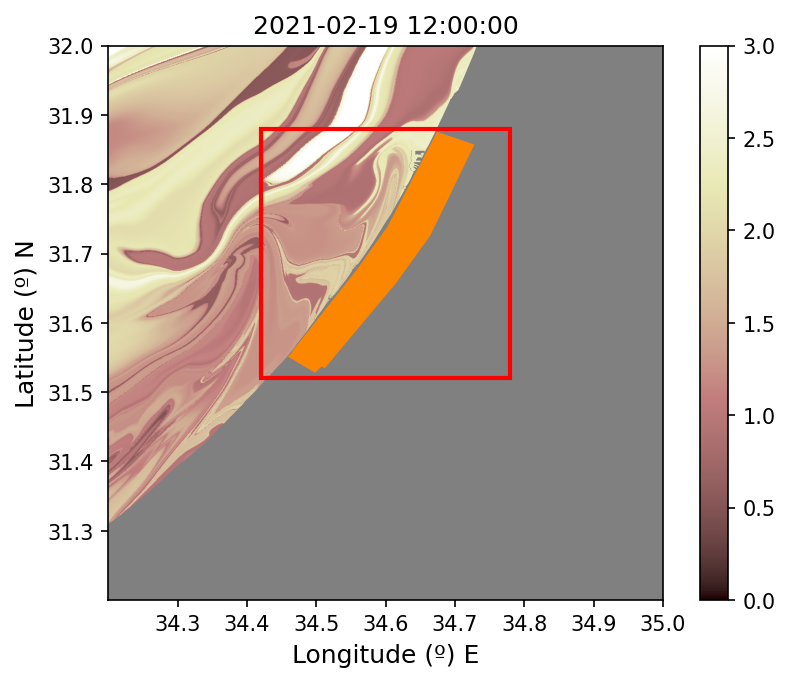} 
    d) \includegraphics[scale=0.4]{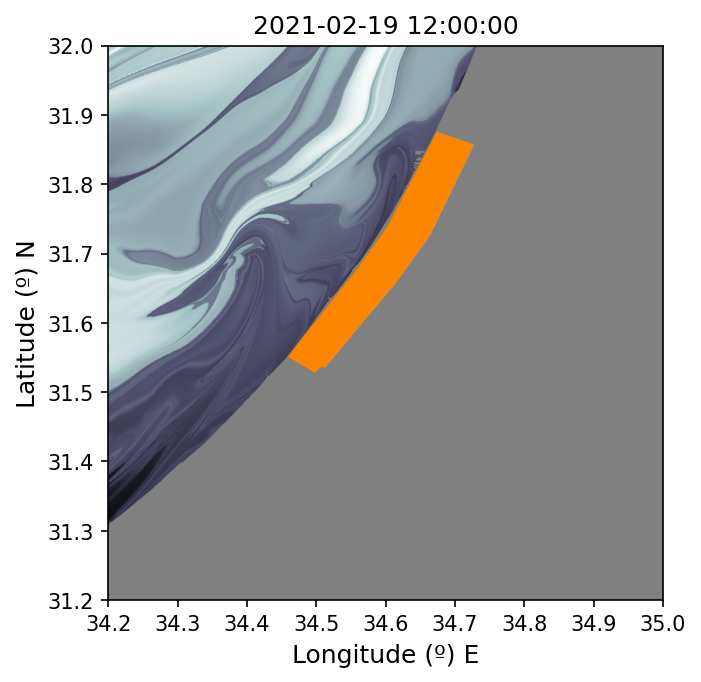}
\end{center}
   \caption{Evaluation on the 19th of February 2021 of  $L_{BUQ}$ using the target, ${\bf x^*}=(34.32^\circ \textrm{E}, 31.35^\circ \textrm{N})$ and of $M^{(b)}$ using $\tau=25$ days; a) $L_{BUQ}$ with the  CMEMS global product; b) $M^{(b)}$ with the  CMEMS global product; c) $L_{BUQ}$ with the  CMEMS Mediterranean product; d) $M^b$ with the  CMEMS Mediterranean product.}
	\label{fig:globmed2}
 \end{figure}
We continue  the discussion considering that the   observation ${\bf x_1}$ is the  orange mark at the Gaza coast (C) on the 19th of February 2021. For it, we consider that the position (A) on the 25th of January 2021 is its backward time target: ${\bf x^*}=(32.319813^\circ \textrm{E},  31.350392^\circ \textrm{N})$. Figure \ref{fig:globmed2} displays the results both for CMEMS global and Mediterranean products. Panels a) and c) display the outputs of  $L_{BUQ}$, and the red box highlights the neighborhood ${\bf X_1}$  used to characterize the performance of the model. Panels b) and d) again confirm the similarities between the unstable manifolds of hyperbolic trajectories in Eq. \eqref{v(t)s} and the structure of   $L_{BUQ}$.

It is remarkable that for both cases examined in figures \ref{fig:globmed1} and \ref{fig:globmed2}, $L_{BUQ}$ takes, in the neighborhood ${\bf X_1}$, considerably larger values for  the CMEMS Mediterranean  product than for the global one. Table \ref{tab} confirms this point by providing the averages in both domains in the column labeled as ``Remote Targets",
suggesting the better performance of the global model for these cases.

\begin{figure}[htb!]
 \begin{center}
   a) \includegraphics[scale=0.4]{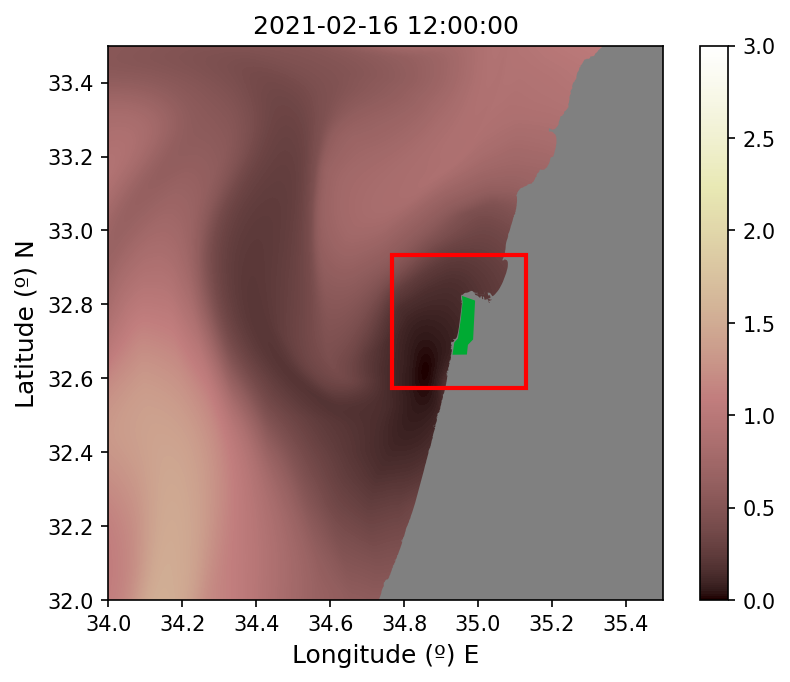} 
    b) \includegraphics[scale=0.4]{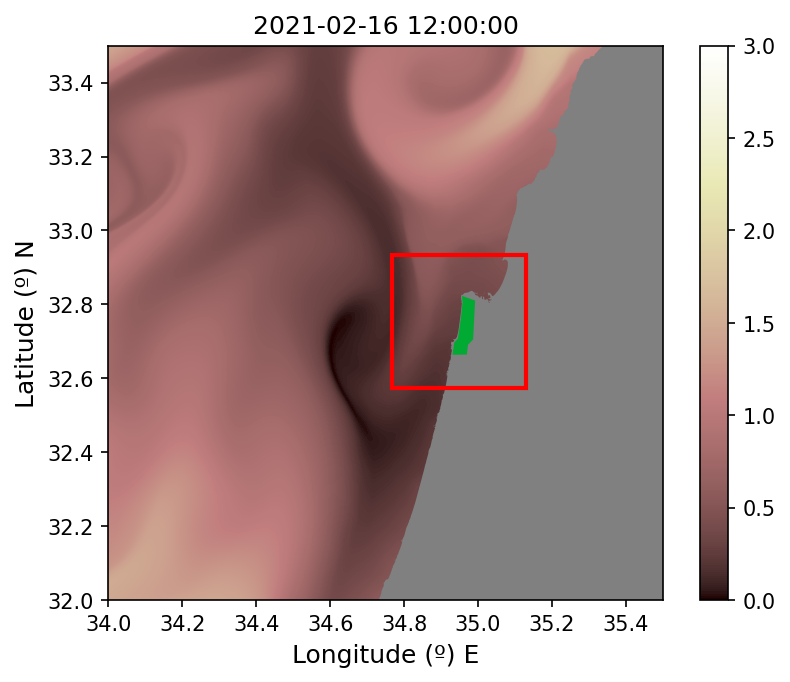} \\
   c) \includegraphics[scale=0.4]{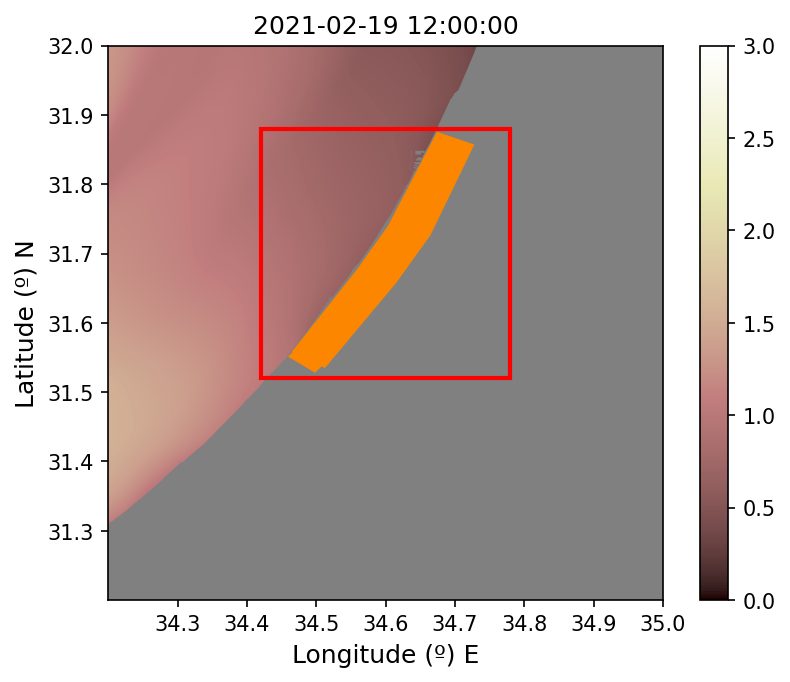} 
    d) \includegraphics[scale=0.4]{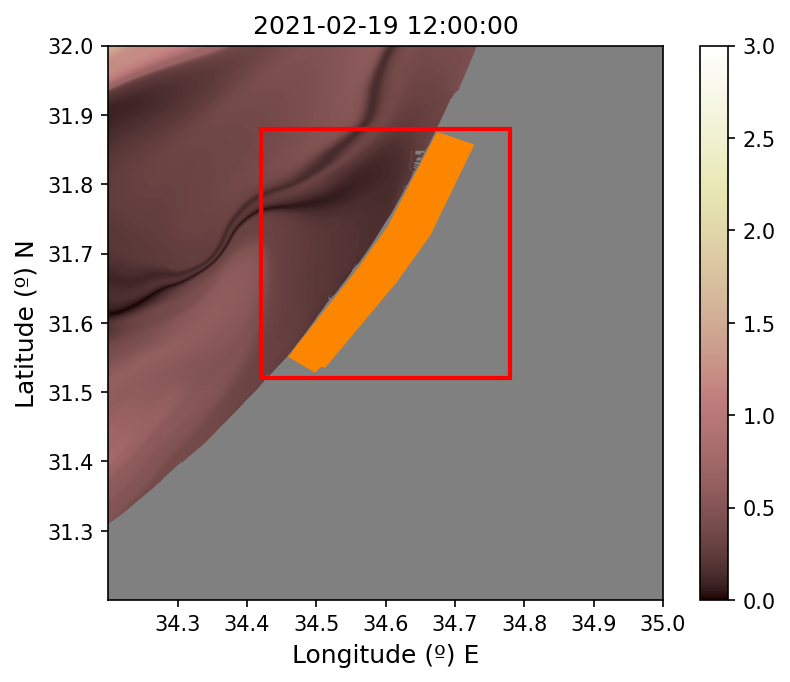}
\end{center}
   \caption{Evaluation $L_{BUQ}$ on the Israel and Gaza coast for the close targets using the global and the Mediterranean data.  $L_{BUQ}$ on 16th of February 2021 using the target, ${\bf x^*}=(34.72^\circ \textrm{E}, 32.59^\circ \textrm{N})$; a) $L_{BUQ}$  on the Israel coast with the  CMEMS global product; b) $M^{(b)}$  on the Israel coast with the  CMEMS Mediterranean product.  $L_{BUQ}$ on 19th of February 2021 using the target, ${\bf x^*}=(34.28^\circ \textrm{E}, 32.59^\circ \textrm{N})$; c) $L_{BUQ}$ on the Gaza coast with the  CMEMS Global product; d) $M^{(b)}$ on the Gaza coast with the  CMEMS Mediterranean product.}
	\label{fig:globmed1short}
 \end{figure}
Figure \ref{fig:med} shows that related to the oil arrivals marked in green (F) and orange (C), there exist  additional  targets ${\bf x^*}$  nearer and closer to the observation ${\bf x_1}$, than those previously discussed. These backward-in-time targets, labeled  (B) and (E), were identified from satellite imagery between the dates of the original sources and the dates of arrival on shore. These spills, after several days have elapsed since their leak, spread out on the ocean surface due to the ocean movement, which stretches and folds them in a chaotic dynamic. Indeed, this description matches well the appearance of targets  (B) and (E), as discussed in \cite{garciasanchez2022}. Now we consider the evaluation of $L_{BUQ}$ using these new ${\bf x^*}$  to assess both models. We start considering the green observation (F)  (${\bf x_1}$) on the Israeli coast. For it, central coordinates of the spill contour on  the 12th of February 2021 are ${\bf x^*}=(34.72008^\circ \textrm{E}, 32.59311^\circ \textrm{N})$. Other  target positions within the extended spill could have been considered, but they do not lead to substantial changes either in the results or in the discussions. Panels a) and b) in figure \ref{fig:globmed1short}
display the evaluation $L_{BUQ}$ for this target, considering the CMEMS global and Mediterranean products respectively. 
Regarding the orange observation (C)  coordinates that correspond to its  backward-in-time target  on  the 12th of February 2021 are ${\bf x^*}=(34.2891^\circ \textrm{E}, 32.59311^\circ \textrm{N})$. As before, given that this is a rather spread contour, other target positions could have been considered, but we do not discuss them as they provide no changes in the conclusions.
For these assumptions  panels c) and d)  
of figure \ref{fig:globmed1short}
display $L_{BUQ}$ considering the CMEMS global and Mediterranean products respectively. 
In these pictures, the red box delimits the boundaries of the neighborhood  ${\bf X_1}$ used to evaluate the model performance. In this case the column ``Close Targets" in Table \ref{tab} shows the results for both models.


A discussion of these results requires highlighting the sizes involved in the different settings.  The distance  between the remote targets and the Gaza (orange) and Israeli (green) shore impact points are, respectively, $\sim$ 215 km and 127 km.
These size ranges 
involve ocean mesoscale structures, which seem to be better represented in the global model. The close targets are  at distances $\sim$ 28 km and 33 km respectively of the Gaza and Israel  arrival points (C) and (F). These sizes   involve ocean submesoscale structures, which should be better represented by the higher resolution model, which is the CMEMS Mediterranean product. Indeed, this is the one that on average performs better for the closer targets, although the global model shows a good performance as well for the Israel case, despite its lower resolution. 


\begin{table}[!htbp]
\begin{tabular}{c|c|c||c|c}
\hline
\multicolumn{1}{c}{}&\multicolumn{2}{c}{Remote Targets}&\multicolumn{2}{c}{Close Targets } \\
\hline
Case & Global & Mediterranean & Global & Mediterranean\\ 
\hline
\hline
Israel &0.2989& 0.9434&0.1642& 0.3375 \\
\hline
Gaza&0.34074&1.5426&0.7810&0.24666\\
\hline 
\end{tabular} 
\medskip
\caption{Averages of $L_{BUQ}$ within  the ${\bf X_1}$ domains highlighted with red boxes.} 
\label{tab}
\end{table}

\section{Conclusions}\label{Conclusions}

This article proposes a new definition  for uncertainty quantification that extends those recently proposed in oceanic contexts in \cite{garciasanchez2020,garciasanchez2022a}. The new definition is referred to as Lagrangian Uncertainty Quantification in Backward time ($L_{BUQ}$), and it is suited to measure if a given transport model consistently describes the source of an observation whose origin is unknown. 

We have found that  the defined quantity $L_{BUQ}$, when evaluated in the neighborhood of an observation with respect to its backward-in-time target,  has a  {\em structure} which has been linked to the unstable invariant manifolds of the hyperbolic trajectories  present in the model vector field. This link  has been rigorously 
proven in a simple example, and numerically verified in a real inspired case: the oil spill accident that affected the Eastern Mediterranean in 2021, which was studied in \cite{garciasanchez2022}. 

The new definition, $L_{BUQ}$, has been exploited to quantify the performance of different CMEMS products to describe the sequence of events reported in  \cite{garciasanchez2022} regarding the oil spill accident in the Eastern Mediterranean in 2021.  In this event, some of the backward-in-time targets are very far from the impact point, and their evolution involves mesoscale structures, which our analysis shows are better represented in the CMEMS global model. On the contrary for targets that are close, involving submesoscale structures, on average the CMEMS Mediterranean product   performs better, although in some cases CMEMS global product also performs well. 

Our results are set within a different framework compared to other inverse uncertainty studies. In those studies, the model typically consists of a dissipative partial differential equation with adjustable parameters, and the uncertainty analysis aims to determine the best-fitting parameters based on observations. However, in our framework, the model comprises a two-dimensional set of volume-preserving (non-dissipative) non-autonomous differential equations, and their exact form is unknown as they rely on approximated velocity fields. 
Many classical inverse problems, such as the inverse heat equation, are ill-posed, meaning that inferring a previous temperature distribution from final data is highly sensitive to changes in the final data. Although our framework differs from such cases, we still observe a high sensitivity to the uncertainty measured around a final observation, as demonstrated in Figures 5-7. In the neighborhood surrounding the final observation, the uncertainty exhibits a complex structure with high oscillations, and we have established a connection with the unstable manifolds of hyperbolic trajectories. Thus, we illustrate how concepts from dynamical systems are linked to uncertainty quantification, whereas a statistical perspective has traditionally been the main focus in this context.  Finally, we have shown  how to implement these concepts for the practical purpose of discriminating the best data set performer on specific environmental applications.

\section*{Acknowledgements}

GGS and AMM acknowledge the support of a CSIC PIE project Ref. 202250E001.  AMM, GGS and MA acknowledge the support from grant PID2021-123348OB-I00 funded by 
MCIN/ AEI /10.13039/501100011033/ and by
FEDER A way to making Europe. MA acknowledges the support from the grant CEX2019-000904-S and IJC2019-040168-I funded by: MCIN/AEI/10.13039/501100011033. AMM, GGS, and MA are active members
of the CSIC Interdisciplinary Thematic Platforms TELEDETECT. SW gratefully acknowledges the support of the William R. Davis '68 Chair at the United States Naval Academy and the support of EPSRC Grant No.~EP/P021123/1.

\section*{Appendix}
Our proof follows the spirit of the work by \cite{garciasanchez2022a,lopesino2017theoretical,lopesino2015lagrangian,garcia2018detection}.  We assume the definition of singular features given there, by considering that these are features of  $L_{BUQ}$  (we consider for simplicity the expression for $L_{BUQ}$  given in Eq. \ref{eq:BUQ}) on which the transversal derivative is not defined. We prove that, for the simple case of  the linear saddle,  the unstable manifold of the hyperbolic fixed point is aligned with singular features of $L_{BUQ}$.

We study the vector field of the autonomous saddle case, where the equations of motion are the following: 

\begin{equation}\label{eq:saddleap}
    \begin{cases} 
    \frac{dx}{dt}  = \lambda x, \\ 
    \frac{dy}{dt}  = -\lambda y,
     \end{cases} \quad \lambda > 0 
\end{equation}

The system has a unique solution for a given  condition $(x_1 , y_1 )$ at time $t=t_1$:

\begin{equation}
    \begin{cases} 
    x(t)  = x_1 e^{\lambda (t-t_1)} \\ 
    y(t)  = y_1 e^{-\lambda (t-t_1)},
    \end{cases} \quad \lambda > 0 
    \label{eq:saddlesolution}
\end{equation}

Notice that the origin $(0,0)$ is a hyperbolic fixed point with stable and unstable manifolds which are the following:

\begin{equation}
    W^s(0,0) = \{ (x,y)\in \mathbb{R}^2 \colon x=0, y\neq 0 \},
    \label{manifold: stable saddle}
\end{equation}
\begin{equation}
    W^u(0,0) = \{ (x,y)\in \mathbb{R}^2 \colon x\neq 0, y= 0 \},
    \label{manifold: unstable saddle}
\end{equation}

Since in this paper we study the backwards uncertainty quantification method we will consider according to  \eqref{eq:BUQ}:

\begin{align}\label{LUQsaddle}
L_{BUQ}({\bf X}_{1}, t_{1}, \tau,p) &= \sum_{i=1}^n\left|x_i(t_{1}-\tau)-x_{i}^*\right|^p \nonumber\\
&= |x_1 e^{\lambda \tilde{t}} - x^*|^p + |y_1 e^{-\lambda \tilde{t}} - y^*|^p \nonumber\\
&= |x_1|^p |\omega-a|^p + |y_1|^p\omega^{-p}|1-b\omega|^p
\end{align}
where $\tilde{t}=t_{1}-\tau-t_1=-\tau<0$, $a = \frac{x^*}{x_1}, b = \frac{y^*}{y_1}$ and $\omega = e^{\lambda \tilde{t}}=e^{-\lambda \tau}$, $p\leq 1$. In order to compute equation (\ref{LUQsaddle}) we will calculate, for simplicity, each of the terms separately. In the first term we notice that the sign of the expression $(\omega-a)$, for $\tau\gg 1$, will be positive when $a<0$ and could be negative when $a>0$.

\begin{itemize}
\item For $a<0$:

\begin{align} \label{dominant positive unstable}
   |\omega-a| ^p &=(\omega -a)^p \\
   &= (-a)^p + p \omega (-a)^{p-1} + \frac{1}{2} (p-1) p \omega^2(-a)^{p-2}
+ \frac{1}{6} (p-2) (p-1) p \omega^3(-a)^{p-3}+O\left(\omega^4\right)\\
&=(-a)^p + O\left(e^{-\lambda \tau}\right), e^{-\lambda \tau} \ll 1 
\end{align}

\item For $a>0$ let be  $t_{L}$ such that for $\tau>t_{L}$, we get that $(\omega-a)<0$:

\begin{align} \label{dominant abs unstable}
      |\omega-a|^p &=(-\omega+a)^p = a^p + p \omega a^{p-1}+\frac{1}{2} (p-1) p \omega^2 a^{p-2}+ \frac{1}{6} (p-2) (p-1) p \omega^3 a^{p-3}+O\left(\omega^4\right)\\
       &= a^p + O\left(\omega\right)
\end{align}

Therefore
\begin{equation}
    |\omega-a|^p = |a|^p+  O\left(e^{-\lambda\tau}\right), \, \omega \ll 1.
    \label{dominant abs unstable}
\end{equation}

Note that $a = x^*/x_1$. Thus
\[
  |x_1|^p  |\omega-a|^p =|x_1|^p|e^{-\lambda \tau}-a|^p =  |x^*|^p+  O\left(|x_1|^p e^{-\lambda \tau}\right), \, e^{-\lambda \tau} \ll 1.
\]

\end{itemize}

For the factor $ |1 - b\omega|^p$  of the second term there exists a $t_L$ such that if $\tau>t_L$ , then $b \omega<<1$ and $(1-b \omega)>0$. This is always the case for $b<0$ and is a plausible assumption for $b>0$, if $b<<\omega^{-1}$. We recall that $b= y^*/y_1$ and that therefore such $t_L$ exists if $y_1 \neq 0$. In this case positiveness is guaranteed for sufficiently large $\tau$, i.e, a Taylor series around $\omega=0$, attained if $\tau\gg1$ and $\tau>t_L$, is performed for the binomial:

\[
(1-b\omega)^p = 1-b p \omega+\frac{1}{2} b^2 (p-1) p \omega^{2}-\frac{1}{6} \omega^{3} \left(b^3 (p-2) (p-1) p\right)+O\left(\omega^{4}\right).
\]
Therefore, 
\[
\frac{1}{\omega^{p}} (1-b\omega)^p 
 = \omega^{-p} - b p \omega^{(1-p)} + \frac{1}{2} b^2 (p-1) p \omega^{(2-p)} - O\left(\omega^{(3-p)}\right).
\]

Finally we get:

\begin{align}
    |y_{1}|^p\omega^{-p}|1-b\omega|^p &=|y_{1}|^{p} |\omega^{-p}-b|^{p}\\
    &=|y_{1}|^p |e^{\lambda\tau p}-b|^{p}\\
    &= |y_{1}|^{p} e ^{\lambda \tau p} + O\left(|y_1|^p b e^{-\lambda \tau (1-p)}\right)
\end{align}

Thus, we can approximate $L_{BUQ}$ as
\begin{equation}
L_{BUQ}(\textbf{X}_{1}, t_{1}, \tau,p) \approx |x^*|^p + |y_1 e^{\lambda\tau }|^{p}  = |x^*|^p + |y_{1}|^{p} e ^{\lambda \tau p}
\label{luqapp1} 
\end{equation}
the dominant term is the term $|y_1|^p e^{\lambda \tau p}$. Therefore the unstable manifold at $y=0$, for `sufficiently large'  $\tau$, is aligned with a singular feature of $L_{BUQ}$, as long as $y_1\neq 0$, and for any $|y_1|> 0$  for a sufficient large $\tau$ satisfying, $\tau>t_L$.

\section*{Declaration of Generative AI and AI-assisted technologies in the writing process}
During the preparation of this work, AMM, MA used Google Translator, Grammarly, and Chat GPT to improve the use of English in writing. 
After using these tools/services, AMM, MA reviewed and edited the content as necessary and take full responsibility for the content of the publication.

\bibliography{SNreac}

\begin{thebibliography}{31}
\expandafter\ifx\csname natexlab\endcsname\relax\def\natexlab#1{#1}\fi
\providecommand{\url}[1]{\texttt{#1}}
\providecommand{\href}[2]{#2}
\providecommand{\path}[1]{#1}
\providecommand{\DOIprefix}{doi:}
\providecommand{\ArXivprefix}{arXiv:}
\providecommand{\URLprefix}{URL: }
\providecommand{\Pubmedprefix}{pmid:}
\providecommand{\doi}[1]{\href{http://dx.doi.org/#1}{\path{#1}}}
\providecommand{\Pubmed}[1]{\href{pmid:#1}{\path{#1}}}
\providecommand{\bibinfo}[2]{#2}
\ifx\xfnm\relax \def\xfnm[#1]{\unskip,\space#1}\fi
\bibitem[{Wu et~al.(2018)Wu, Kozlowski, Meidani, and Shirvan}]{wu2018inverse}
\bibinfo{author}{X.~Wu}, \bibinfo{author}{T.~Kozlowski},
  \bibinfo{author}{H.~Meidani}, \bibinfo{author}{K.~Shirvan},
\newblock \bibinfo{title}{Inverse uncertainty quantification using the modular
  bayesian approach based on gaussian process, part 2: Application to trace},
\newblock \bibinfo{journal}{Nuclear Engineering and Design}
  \bibinfo{volume}{335} (\bibinfo{year}{2018}) \bibinfo{pages}{417--431}.
\bibitem[{Domitr et~al.(2022)Domitr, W{\l}ostowski, Laskowski, and
  Jurkowski}]{domitr2022comparison}
\bibinfo{author}{P.~Domitr}, \bibinfo{author}{M.~W{\l}ostowski},
  \bibinfo{author}{R.~Laskowski}, \bibinfo{author}{R.~Jurkowski},
\newblock \bibinfo{title}{Comparison of inverse uncertainty quantification
  methods for critical flow test},
\newblock \bibinfo{journal}{Energy}  (\bibinfo{year}{2022})
  \bibinfo{pages}{125640}.
\bibitem[{de~Cr{\`e}cy(1996)}]{de1996determination}
\bibinfo{author}{A.~de~Cr{\`e}cy}, \bibinfo{title}{Determination of the
  uncertainties of the constitutive relationships in the CALTHARE 2 code},
  \bibinfo{type}{Technical Report}, American Society of Mechanical Engineers,
  New York, NY (United States), \bibinfo{year}{1996}.
\bibitem[{Petruzzi(2019)}]{petruzzi2019casualidad}
\bibinfo{author}{A.~Petruzzi},
\newblock \bibinfo{title}{The casualidad method for uncertainty evaluation of
  best-estimate system thermal-hydraulic calculations},
\newblock \bibinfo{journal}{Nuclear Technology} \bibinfo{volume}{205}
  (\bibinfo{year}{2019}) \bibinfo{pages}{1554--1566}.
\bibitem[{Garc{\'i}a-S{\'a}nchez et~al.(2022)Garc{\'i}a-S{\'a}nchez, Mancho,
  and Wiggins}]{guillermo2022}
\bibinfo{author}{G.~Garc{\'i}a-S{\'a}nchez}, \bibinfo{author}{A.~Mancho},
  \bibinfo{author}{S.~Wiggins},
\newblock \bibinfo{title}{A bridge between invariant dynamical structures and
  uncertainty quantification},
\newblock \bibinfo{journal}{Commun Nonlinear Sci Numer Simulat}
  \bibinfo{volume}{104} (\bibinfo{year}{2022}) \bibinfo{pages}{106016}.
  \DOIprefix\doi{10.1016/j.cnsns.2021.106016}.
\bibitem[{Garc{\'\i}a-S{\'a}nchez et~al.(2021)Garc{\'\i}a-S{\'a}nchez, Mancho,
  Ramos, Coca, P{\'e}rez-G{\'o}mez, {\'A}lvarez-Fanjul, Sotillo,
  Garc{\'\i}a-Le{\'o}n, Garc{\'\i}a-Garrido, and Wiggins}]{garciasanchez2020}
\bibinfo{author}{G.~Garc{\'\i}a-S{\'a}nchez}, \bibinfo{author}{A.~M. Mancho},
  \bibinfo{author}{A.~G. Ramos}, \bibinfo{author}{J.~Coca},
  \bibinfo{author}{B.~P{\'e}rez-G{\'o}mez},
  \bibinfo{author}{E.~{\'A}lvarez-Fanjul}, \bibinfo{author}{M.~G. Sotillo},
  \bibinfo{author}{M.~Garc{\'\i}a-Le{\'o}n}, \bibinfo{author}{V.~J.
  Garc{\'\i}a-Garrido}, \bibinfo{author}{S.~Wiggins},
\newblock \bibinfo{title}{Very high resolution tools for the monitoring and
  assessment of environmental hazards in coastal areas},
\newblock \bibinfo{journal}{Frontiers in Marine Science} \bibinfo{volume}{7}
  (\bibinfo{year}{2021}).
\bibitem[{Garc{\'\i}a-S{\'a}nchez et~al.(2022)Garc{\'\i}a-S{\'a}nchez, Mancho,
  G., J., and Wiggins}]{garciasanchez2022}
\bibinfo{author}{G.~Garc{\'\i}a-S{\'a}nchez}, \bibinfo{author}{A.~M. Mancho},
  \bibinfo{author}{R.~A. G.}, \bibinfo{author}{C.~J.},
  \bibinfo{author}{S.~Wiggins},
\newblock \bibinfo{title}{Structured pathways in the turbulence organizing
  recent oil spill events in the eastern mediterranean},
\newblock \bibinfo{journal}{Scientific Reports} \bibinfo{volume}{12}
  (\bibinfo{year}{2022}) \bibinfo{pages}{3662}.
\bibitem[{Brushett et~al.(2011)Brushett, King, and
  Lemckert}]{brushett2011evaluation}
\bibinfo{author}{B.~A. Brushett}, \bibinfo{author}{B.~A. King},
  \bibinfo{author}{C.~Lemckert},
\newblock \bibinfo{title}{Evaluation of met-ocean forecast data effectiveness
  for tracking drifters deployed during operational oil spill response in
  australian waters},
\newblock \bibinfo{journal}{Journal of Coastal Research}
  (\bibinfo{year}{2011}) \bibinfo{pages}{991--994}.
\bibitem[{Zhang et~al.(2020)Zhang, Cheng, Zhang, Wu, Li, Liu, Chu, Xia, Min,
  Zuo et~al.}]{zhang2020evaluation}
\bibinfo{author}{X.~Zhang}, \bibinfo{author}{L.~Cheng},
  \bibinfo{author}{F.~Zhang}, \bibinfo{author}{J.~Wu}, \bibinfo{author}{S.~Li},
  \bibinfo{author}{J.~Liu}, \bibinfo{author}{S.~Chu}, \bibinfo{author}{N.~Xia},
  \bibinfo{author}{K.~Min}, \bibinfo{author}{X.~Zuo}, et~al.,
\newblock \bibinfo{title}{Evaluation of multi-source forcing datasets for drift
  trajectory prediction using lagrangian models in the south china sea},
\newblock \bibinfo{journal}{Applied Ocean Research} \bibinfo{volume}{104}
  (\bibinfo{year}{2020}) \bibinfo{pages}{102395}.
\bibitem[{{Wikipedia}(2021)}]{WP}
\bibinfo{author}{{Wikipedia}}, \bibinfo{title}{2021 Mediterranean Oil Spill.},
  \bibinfo{organization}{Wikipedia},
  \bibinfo{address}{https://en.wikipedia.org/wiki/2021\_Mediterranean\_oil\_spill},
  \bibinfo{year}{2021}.
\bibitem[{{Z. Rinat and A. Ben Zikri}(2021)}]{haar}
\bibinfo{author}{{Z. Rinat and A. Ben Zikri}}, \bibinfo{title}{Oil Spill Off
  Israel's Coast Is Its Worst Maritime Pollution in Decades, and Cleanup 'Could
  Take Years'}, \bibinfo{organization}{Haaretz},
  \bibinfo{address}{https://www.haaretz.com/israel-news/.premium-oil-spill-off-israel-s-coast-is-its-worst-maritime-pollution-in-decades-1.9553528},
  \bibinfo{year}{2021}.
\bibitem[{Tercatin(2021)}]{jp1}
\bibinfo{author}{R.~Tercatin}, \bibinfo{title}{Damage to Israeli marine
  environment from tar spill extreme, experts say},
  \bibinfo{organization}{Jerusalem Post},
  \bibinfo{address}{https://www.jpost.com/israel-news/damage-to-israeli-marine-environment-from-tar-spill-extreme-experts-say-659716},
  \bibinfo{year}{2021}.
\bibitem[{Kaplan-Zantopp(2021)}]{jp2}
\bibinfo{author}{M.~Kaplan-Zantopp}, \bibinfo{title}{Israel oil spill: How did
  it happen and what will we do going forward?},
  \bibinfo{organization}{Jerusalem Post},
  \bibinfo{address}{https://www.jpost.com/israel-news/israel-oil-spill-how-did-it-happen-and-what-will-we-do-going-forward-660242},
  \bibinfo{year}{2021}.
\bibitem[{{T. Joffre and R. Tercatin }(2021)}]{jp3}
\bibinfo{author}{{T. Joffre and R. Tercatin }}, \bibinfo{title}{Israel
  investigates tar spill calamity, places inquiry under gag order},
  \bibinfo{organization}{Jerusalem Post},
  \bibinfo{address}{https://www.jpost.com/israel-news/israel-oil-spill-disaster-investigation-details-placed-under-censor-659765},
  \bibinfo{year}{2021}.
\bibitem[{{AL JAZEERA AND NEWS AGENCIES}(2021)}]{ay1}
\bibinfo{author}{{AL JAZEERA AND NEWS AGENCIES}}, \bibinfo{title}{Lebanon
  begins cleaning beaches after oil spill}, \bibinfo{organization}{Aljazeera},
  \bibinfo{address}{https://www.aljazeera.com/news/2021/2/27/lebanon-begins-cleaning-beaches-after-oil-spill},
  \bibinfo{year}{2021}.
\bibitem[{{Rutgers Staff}(2021)}]{rut}
\bibinfo{author}{{Rutgers Staff}}, \bibinfo{title}{Oil spill off Israel reaches
  south Lebanese beaches}, \bibinfo{organization}{Rutgers},
  \bibinfo{address}{https://www.reuters.com/article/us-israel-environment-oil-spill-lebanon-idUSKBN2AM19V},
  \bibinfo{year}{2021}.
\bibitem[{Polidura(2021)}]{ata}
\bibinfo{author}{A.~Polidura}, \bibinfo{title}{Israel's oil spill now affects
  entire Lebanese coastline}, \bibinfo{organization}{Atalayar},
  \bibinfo{address}{https://atalayar.com/en/content/israels-oil-spill-now-affects-entire-lebanese-coastline},
  \bibinfo{year}{2021}.
\bibitem[{Garc{\'i}a-Garrido et~al.(2018)Garc{\'i}a-Garrido, Curbelo, Mancho,
  Wiggins, and Mechoso}]{gg2018}
\bibinfo{author}{V.~J. Garc{\'i}a-Garrido}, \bibinfo{author}{J.~Curbelo},
  \bibinfo{author}{A.~M. Mancho}, \bibinfo{author}{S.~Wiggins},
  \bibinfo{author}{C.~R. Mechoso},
\newblock \bibinfo{title}{The application of lagrangian descriptors to 3d
  vector fields},
\newblock \bibinfo{journal}{Regul Chaotic Dyn} \bibinfo{volume}{23}
  (\bibinfo{year}{2018}) \bibinfo{pages}{551--568}.
  \DOIprefix\doi{10.1134/S1560354718050052}.
\bibitem[{Haller(2002)}]{haller2002}
\bibinfo{author}{G.~Haller},
\newblock \bibinfo{title}{Lagrangian coherent structures from approximate
  velocity data.},
\newblock \bibinfo{journal}{Phys. Fluids} \bibinfo{volume}{14}
  (\bibinfo{year}{2002}) \bibinfo{pages}{1851–1861}.
\bibitem[{Kuznetsov et~al.(2002)Kuznetsov, Toner, Kirwan, Jones, Kantha, and
  Choi}]{kuznetsov2002}
\bibinfo{author}{L.~Kuznetsov}, \bibinfo{author}{M.~Toner},
  \bibinfo{author}{A.~D. Kirwan}, \bibinfo{author}{C.~K. R.~T. Jones},
  \bibinfo{author}{L.~H. Kantha}, \bibinfo{author}{J.~Choi},
\newblock \bibinfo{title}{{The Loop Current and adjacent rings delineated by
  Lagrangian analysis of the near-surface flow}},
\newblock \bibinfo{journal}{J. Mar. Res.} \bibinfo{volume}{60}
  (\bibinfo{year}{2002}) \bibinfo{pages}{405–429}.
\bibitem[{Shadden et~al.(2009)Shadden, Lekien, Paduan, Chavez, and
  Marsden}]{shadden2009}
\bibinfo{author}{S.~C. Shadden}, \bibinfo{author}{F.~Lekien},
  \bibinfo{author}{J.~D. Paduan}, \bibinfo{author}{F.~P. Chavez},
  \bibinfo{author}{J.~E. Marsden},
\newblock \bibinfo{title}{{The correlation between surface drifters and
  coherent structures based on high-frequency radar data in Monterey Bay}},
\newblock \bibinfo{journal}{Deep Sea Res. II} \bibinfo{volume}{56}
  (\bibinfo{year}{2009}) \bibinfo{pages}{161--172}.
\bibitem[{Haza et~al.(2007)Haza, Griffa, Martin, Molcard, {\"Ozg\"okmen}, Poje,
  Barbanti, Book, Poulain, Rixen, and Zanasca}]{haza2007}
\bibinfo{author}{A.~C. Haza}, \bibinfo{author}{A.~Griffa},
  \bibinfo{author}{P.~Martin}, \bibinfo{author}{A.~Molcard},
  \bibinfo{author}{T.~M. {\"Ozg\"okmen}}, \bibinfo{author}{A.~Poje},
  \bibinfo{author}{R.~Barbanti}, \bibinfo{author}{J.~Book},
  \bibinfo{author}{P.~Poulain}, \bibinfo{author}{M.~Rixen},
  \bibinfo{author}{P.~Zanasca},
\newblock \bibinfo{title}{{Model-based directed drifter launches in the
  Adriatic Sea: Results from the DART experiment}},
\newblock \bibinfo{journal}{Geophys. Res. Lett} \bibinfo{volume}{34}
  (\bibinfo{year}{2007}) \bibinfo{pages}{L10605}.
\bibitem[{Haza et~al.(2010)Haza, {\"Ozg\"okmen}, Griffa, Molcard, Poulain, and
  Peggion}]{haza2010}
\bibinfo{author}{A.~C. Haza}, \bibinfo{author}{T.~{\"Ozg\"okmen}},
  \bibinfo{author}{A.~Griffa}, \bibinfo{author}{A.~Molcard},
  \bibinfo{author}{P.~M. Poulain}, \bibinfo{author}{G.~Peggion},
\newblock \bibinfo{title}{{Transport properties in small-scale coastal flows:
  relative dispersion from VHF radar measurements in the Gulf of La Spezia}},
\newblock \bibinfo{journal}{Ocean Dynam.} \bibinfo{volume}{60}
  (\bibinfo{year}{2010}) \bibinfo{pages}{861–882}.
\bibitem[{Mendoza et~al.(2014)Mendoza, Mancho, and
  Wiggins}]{mendoza2014lagrangian}
\bibinfo{author}{C.~Mendoza}, \bibinfo{author}{A.~Mancho},
  \bibinfo{author}{S.~Wiggins},
\newblock \bibinfo{title}{Lagrangian descriptors and the assessment of the
  predictive capacity of oceanic data sets},
\newblock \bibinfo{journal}{Nonlinear Processes in Geophysics}
  \bibinfo{volume}{21} (\bibinfo{year}{2014}) \bibinfo{pages}{677--689}.
\bibitem[{Garcia-Garrido et~al.(2015)Garcia-Garrido, Mancho, Wiggins, and
  Mendoza}]{gg2015}
\bibinfo{author}{V.~J. Garcia-Garrido}, \bibinfo{author}{A.~M. Mancho},
  \bibinfo{author}{S.~Wiggins}, \bibinfo{author}{C.~Mendoza},
\newblock \bibinfo{title}{{A dynamical systems approach to the surface search
  for debris associated with the disappearance of flight MH370.}},
\newblock \bibinfo{journal}{Nonlin. Processes Geophys.} \bibinfo{volume}{22}
  (\bibinfo{year}{2015}) \bibinfo{pages}{701--712}.
  \DOIprefix\doi{10.5194/npg-22-701-2015}.
\bibitem[{Mendoza and Mancho(2010)}]{mendoza2010}
\bibinfo{author}{C.~Mendoza}, \bibinfo{author}{A.~M. Mancho},
\newblock \bibinfo{title}{Hidden geometry of ocean flows},
\newblock \bibinfo{journal}{Phys Rev Lett} \bibinfo{volume}{105}
  (\bibinfo{year}{2010}) \bibinfo{pages}{038501}.
  \DOIprefix\doi{10.1103/PhysRevLett.105.038501}.
\bibitem[{Mancho et~al.(2013)Mancho, Wiggins, Curbelo, and
  Mendoza}]{mancho2013lagrangian}
\bibinfo{author}{A.~M. Mancho}, \bibinfo{author}{S.~Wiggins},
  \bibinfo{author}{J.~Curbelo}, \bibinfo{author}{C.~Mendoza},
\newblock \bibinfo{title}{Lagrangian descriptors: A method for revealing phase
  space structures of general time dependent dynamical systems},
\newblock \bibinfo{journal}{Commun. Nonlinear Sci. Numer. Simul.}
  \bibinfo{volume}{18} (\bibinfo{year}{2013}) \bibinfo{pages}{3530--3557}.
  \DOIprefix\doi{https://doi.org/10.1016/j.cnsns.2013.05.002}.
\bibitem[{Garc{\'\i}a-S{\'a}nchez et~al.(2022)Garc{\'\i}a-S{\'a}nchez, Mancho,
  and Wiggins}]{garciasanchez2022a}
\bibinfo{author}{G.~Garc{\'\i}a-S{\'a}nchez}, \bibinfo{author}{A.~M. Mancho},
  \bibinfo{author}{S.~Wiggins},
\newblock \bibinfo{title}{A bridge between invariant dynamical structures and
  uncertainty quantification},
\newblock \bibinfo{journal}{Commun. Nonlinear Sci. Numer. Simul.}
  \bibinfo{volume}{104} (\bibinfo{year}{2022}).
\bibitem[{Lopesino et~al.(2017)Lopesino, Balibrea-Iniesta, Garc{\'\i}a-Garrido,
  Wiggins, and Mancho}]{lopesino2017theoretical}
\bibinfo{author}{C.~Lopesino}, \bibinfo{author}{F.~Balibrea-Iniesta},
  \bibinfo{author}{V.~J. Garc{\'\i}a-Garrido}, \bibinfo{author}{S.~Wiggins},
  \bibinfo{author}{A.~M. Mancho},
\newblock \bibinfo{title}{A theoretical framework for lagrangian descriptors},
\newblock \bibinfo{journal}{International Journal of Bifurcation and Chaos}
  \bibinfo{volume}{27} (\bibinfo{year}{2017}) \bibinfo{pages}{1730001}.
\bibitem[{Lopesino et~al.(2015)Lopesino, Balibrea, Wiggins, and
  Mancho}]{lopesino2015lagrangian}
\bibinfo{author}{C.~Lopesino}, \bibinfo{author}{F.~Balibrea},
  \bibinfo{author}{S.~Wiggins}, \bibinfo{author}{A.~M. Mancho},
\newblock \bibinfo{title}{Lagrangian descriptors for two dimensional, area
  preserving, autonomous and nonautonomous maps},
\newblock \bibinfo{journal}{Communications in Nonlinear Science and Numerical
  Simulation} \bibinfo{volume}{27} (\bibinfo{year}{2015})
  \bibinfo{pages}{40--51}.
\bibitem[{Garc{\'\i}a-Garrido et~al.(2018)Garc{\'\i}a-Garrido,
  Balibrea-Iniesta, Wiggins, Mancho, and Lopesino}]{garcia2018detection}
\bibinfo{author}{V.~J. Garc{\'\i}a-Garrido},
  \bibinfo{author}{F.~Balibrea-Iniesta}, \bibinfo{author}{S.~Wiggins},
  \bibinfo{author}{A.~M. Mancho}, \bibinfo{author}{C.~Lopesino},
\newblock \bibinfo{title}{Detection of phase space structures of the cat map
  with lagrangian descriptors},
\newblock \bibinfo{journal}{Regular and Chaotic Dynamics} \bibinfo{volume}{23}
  (\bibinfo{year}{2018}) \bibinfo{pages}{751--766}.

\end{thebibliography}

\end{document}